\documentclass[12pt]{article} \usepackage{dina4p} \usepackage{my}
\usepackage{epsfig,amsmath} \usepackage{graphics}
\usepackage{amsfonts}
\newcommand{\CCFM}{CCFMa,CCFMb,CCFMc,CCFMd}

\newcommand{\JETSETMC}{Jetsetc,Pythia61}
\def\JETSET{{\sc Jetset}/{\sc Pythia}}

\newcommand{\RAPGAPMC}{RAPGAP,RAPGAP206}

\def\RAPGAP{{\sc Rapgap}}
\def\lsim{\mathrel{\rlap{\lower4pt\hbox{\hskip1pt$\sim$}}
    \raise1pt\hbox{$<$}}}                % less than or approx. symbol
\def\gsim{\mathrel{\rlap{\lower4pt\hbox{\hskip1pt$\sim$}}
    \raise1pt\hbox{$>$}}}                % greater than or approx. symbol
 
\renewcommand{\thefootnote}{\fnsymbol{footnote}} 
\newcommand{\be}{\begin{equation}}   
\newcommand{\ee}{\end{equation}}   
\newcommand{\beqn}{\begin{eqnarray}}   
\newcommand{\eeqn}{\end{eqnarray}}
\begin{document}
\begin{flushright}
DESY 01-116\\
LUNFD6/(NFFL--7204) 2001 \\
hep-ph/0204269 \\
\end{flushright}
\vspace*{10mm}
\begin{center}  \begin{Large} \begin{bf}
 Massive c${\bf\bar c}$g - Calculation in Diffractive DIS 
 and Diffractive $D^*$ - Production at HERA
  \end{bf}  \end{Large} \\
  \vspace*{5mm}
  \begin{large}
      J.~Bartels$^1$, H.~Jung$^2$, A.~Kyrieleis$^1$\\   
        $^1${\sl II.\ Institut f\"ur Theoretische Physik,
        Universit\"at Hamburg, Germany
          \footnote[1]{Supported by the 
    TMR Network ``QCD and Deep Structure of Elementary Particles''}} \\
          $^2${\sl Department of Physics, Lund University, Sweden}
  \end{large}% 
\begin{quotation}
\noindent
{\bf Abstract:}  
We calculate the cross section for $c\bar{c}g$-production in
diffractive DIS with finite quark masses at zero momentum transfer $t$.
The calculation is done in the leading log(1/$x_{\mathbb P}$)
\mbox{approximation} and is valid in the region of high diffractive masses $M$
(small $\beta$). We apply our cross section formula including
both $c{\bar c}$- and $c\bar{c}g$
in a Monte Carlo simulation to diffractive $D^{*\pm}$ meson
production at HERA.
We compare our predictions to results of H1 
using three parameterizations for the unintegrated gluon density.
\end{quotation}
\end{center}
 \renewcommand{\thefootnote}{\arabic{footnote}} 
 \setcounter{footnote}{0} 
%%%%%%%%%%%%%%%%%%%%%%%%%%%%%%%%%%%%%%%%%%%%%%%%%%%%%%%%%%%%%%%%%%%%%%%%%%%%%%%%%%%%%%%%%%%%%%
\section{Introduction}
%%%%%%%%%%%%%%%%%%%%%%%%%%%%%%%%%%%%%%%%%%%%%%%%%%%%%%%%%%%%%%%%%%%%%%%%%%%%%%%%%%%%%%%%%%%%%%
In the process of diffractive deep inelastic scattering,
$\gamma^*+p\rightarrow p+X$, one can separate perturbative and 
non-perturbative contributions by filtering out particular diffractive final 
states. Examples of diffractive states which are perturbatively calculable 
are longitudinal vector particles or final states which
consist of hard jets (and no soft remnant). In the latter 
case the hard scale which allows the use of pQCD is provided by the
large transverse momenta of the jets, and the Pomeron exchange is modeled
by the unintegrated gluon density. Another particularly interesting 
example is diffractive charm production, since the charm quark mass justifies 
pQCD, even for not so large transverse momenta of the outgoing quarks and 
gluons. Calculations for the diffractive production of massless 
open $q\bar q$ states and of massless $q\bar qg$ states have been 
reported in \cite{qq,nikol,gots} and in \cite{qqg}, respectively, and a comparison
of diffractive two-jet and three-jet events observed at HERA 
with these calculations has been presented in \cite{h1_diff_dijets}. 
Final states with finite
quark masses have been calculated, so far, only for $q\bar{q}$ production 
\cite{qqm} which is expected to be the dominant final state in the region of
small diffractive masses (large $\beta$). 
However, as there are recent measurements 
of diffractive $D^{*\pm}$ - production from the H1~\cite{h1_diff_charm} and 
ZEUS~\cite{zeus_diffcharm_ICHEP00}  collaborations at HERA, 
which extend into the small-$\beta$-region, gluon radiation
can certainly not be neglected, and a full perturbative calculation of  
$q\bar{q}g$ is needed.
\par
In this article we report
on a calculation of massive $q\bar{q}g$-production in diffractive deep inelastic scattering,
and we present a comparison of our cross section formula with the measurements of
H1~\cite{h1_diff_charm}.  
%%%%%%%%%%%%%%%%%%%%%%%%%%%%%%%%%%%%%%%%%%%%%%%%%%%%%%%%%%%%%%%%%%%%%%%%%%%%%%%%%%%%%%%%%%%%%%
\section{Calculation of massive $q \bar{q} g$ production}
\begin{figure}[htb] 
 \begin{center}
   \input{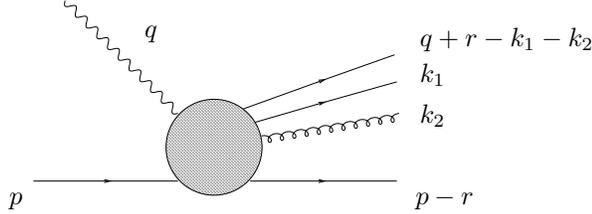} 
   \caption{\label{fig:notat}Kinematics of diffractive $q{\bar q}g$ production}
 \end{center}
\end{figure}

%%%%%%%%%%%%%%%%%%%%%%%%%%%%%%%%%%%%%%%%%%%%%%%%%%%%%%%%%%%%%%%%%%%%%%%%%%%%%%%%%%%%%%%%%%%%%%
We will follow the study of massless $q\bar{q}g$-production 
presented in~\cite{qqg}. In particular, we again work in the leading-log $M^2$ 
approximation, which limits the applicability of our results to the small 
$\beta$-region. Fig.~\ref{fig:notat} shows the notations of the process. As
in~\cite{qqg} we restrict ourselves to zero momentum transfer, $t=r^2=0$. 
As usually, $Q^ 2$ denotes the virtuality of the photon, $\sqrt{W^2}$ the 
energy of the photon proton system, $M$ the mass of the diffractive system, 
and $x=Q^2/(Q^2+W^2)$, $y=pq/pl$ are the Bjorken scaling variables 
(with $l$ being the momentum of the incoming electron). The variable $\beta$ is 
defined as $\beta=Q^2/(Q^2+M^2)$, and it is convenient to introduce 
the momentum fraction of the Pomeron by 
$x_\mathbb P=(Q^2+M^2)/(Q^2+W^2)$. 
\par
We restrict our calculation to the region (leading-log $M^2$ approximation):
\beqn
 Q^2 \ll M^2 \ll W^2. 
\eeqn
%%%%%%%%%%%
%%%%%%%%%%
We use Sudakov variables $k_i=\alpha_i q' + \beta_i p + k_{i\;t}$ (with
$q'=q+xp$, $k_{i\;t}^2=- {\bf k}_i^2)$, and we express the phase space in 
terms of 
$y$, $Q^2$, $M^2$, $m^2$, $t$, ${\bf k}_1^2$, ${\bf k}_2^2$ with 
$m^2=m_{qq}^2 + {\bf k}_2^2$ ($m_{qq}$ denotes the invariant mass of the 
$q\bar{q}$-subsystem). We obtain the 
following result:
$$\frac{d\sigma_D^{e^-p}}{dy dQ^2 dM^2 dm^2 d^2{\bf k}_1 d^2{\bf k}_2 
dt}_{|t=0}=\frac{\alpha_{em}}{yQ^2\pi} \cdot$$ 
\begin{eqnarray}
\lefteqn{ \cdot\left[ \frac{1+(1-y)^2}{2} \frac{d\sigma^{\gamma^*p}_{D,T+}}{dM^2 dm^2 d^2{\bf k}_1 d^2{\bf k}_2 dt_{|t=0}}-2(1-y) \frac{d\sigma^{\gamma^*p}_{D,T-}}{dM^2 dm^2 d^2{\bf k}_1 d^2{\bf k}_2 dt_{|t=0}}  \right. } \nonumber \\
&&\left.{} +(1-y) \frac{d\sigma^{\gamma^*p}_{D,L}}{dM^2 dm^2 d^2{\bf k}_1 d^2{\bf k}_2 dt_{|t=0}} 
+(2-y)\sqrt{1-y} \frac{d\sigma^{\gamma^*p}_{D,I}}{dM^2 dm^2 d^2{\bf k}_1 d^2{\bf k}_2 dt_{|t=0}} \right] \,,
\label{1}
\end{eqnarray}
The differential cross sections of $\gamma^* p \to c c g + p $ are given by: 
\begin{eqnarray}
 \frac{d\sigma^{\gamma^*p}_{D,T+}}{dM^2 dm^2 d^2{\bf k}_1 d^2{\bf k}_2 dt_{|t=0}}&=& \frac{9}{128\pi}\frac{1}{\sqrt{S}(M^2-m^2)m^2} e^2_c \alpha_{em} \alpha^3_s \alpha_1(1-\alpha_1)\cdot \nonumber \\
&& \hspace{1cm} \cdot \left[ \left(\alpha_1^2+(1-\alpha_1)^2 \right)
M_{il} M'_{il}+m_q^2M_lM'_l\right]
\label{2} 
\end{eqnarray}
\begin{eqnarray}
 \frac{d\sigma^{\gamma^*p}_{D,T-}}{dM^2 dm^2 d^2{\bf k}_1 d^2{\bf k}_2 dt_{|t=0}}&=& \frac{9}{128\pi}\frac{1}{\sqrt{S}(M^2-m^2)m^2} e^2_c \alpha_{em} \alpha^3_s \alpha_1^2(1-\alpha_1)^2 \cdot \nonumber \\
&& \hspace{1cm} \cdot \left[ M_{1l} M'_{1l} - M_{2l}M'_{2l} \right]
\label{3} 
\end{eqnarray}
\begin{eqnarray}
 \frac{d\sigma^{\gamma^*p}_{D,L}}{dM^2 dm^2 d^2{\bf k}_1 d^2{\bf k}_2 dt_{|t=0}}&=& \frac{9}{128\pi}\frac{1}{\sqrt{S}(M^2-m^2)m^2} e^2_c \alpha_{em} \alpha^3_s 4\alpha_1^3(1-\alpha_1)^3 Q^2 M_l M'_l
\label{4}
\end{eqnarray}
\begin{eqnarray}
 \frac{d\sigma^{\gamma^*p}_{D,I}}{dM^2 dm^2 d^2{\bf k}_1 d^2{\bf k}_2 dt_{|t=0}}&=& \frac{9}{128\pi}\frac{1}{\sqrt{S}(M^2-m^2)m^2} e^2_c \alpha_{em} \alpha^3_s \alpha_1^2(1-\alpha_1)^2 (1-2\alpha_1) \cdot \nonumber \\
&& \hspace{1cm} \cdot  \sqrt{Q^2} \left[M_{1l} M'_l + M_l M'_{1l} \right]
\label{5} 
\end{eqnarray}
with
\begin{equation}
S=\left( 1+ \frac{{\bf k}^2_1}{m^2} - \frac{({\bf k}_1 +{\bf
k}_2)^2}{m^2} \right)^2 -4\frac{({\bf k}_1^2 + m_q^2)}{m^2}, 
\end{equation}
\begin{equation}
M_{il}=\int \frac{d^2{\bf l}}{\pi{\bf l}^2} {\cal F}(x_{\mathbb P},
{\bf l}^2)T_{il},
\label{9}
\end{equation}
and
\[
T_{il}=\left( \frac{{\bf l}+{\bf k}_1+{\bf k}_2}{D({\bf l}+{\bf
k}_1+{\bf k}_2)} +
\frac{{\bf k}_1+{\bf k}_2}{D({\bf k}_1+{\bf k}_2 )} - \frac{{\bf
k}_1-{\bf l}}{D({\bf k}_1-{\bf l})} - \frac{{\bf k}_1}{D({\bf k}_1)} \right)_i
\left(\frac{{\bf l}+{\bf k}_2}{({\bf l}+{\bf k}_2)^2} - \frac{{\bf k}_2}{{\bf k}_2^2} \right)_l 
\]
\begin{equation}
\hspace{3cm}{}+({\bf l} \to -{\bf l})
\label{Til}
\end{equation}
\[
T_{l}=\left( \frac{1}{D({\bf l}+{\bf k}_1+{\bf k}_2)} +
\frac{1}{D({\bf k}_1+{\bf k}_2)} - \frac{1}{D({\bf k}_1-{\bf l})} -
\frac{1}{D({\bf k}_1)} \right)
\left(\frac{{\bf l}+{\bf k}_2}{({\bf l}+{\bf k}_2)^2} - \frac{{\bf k}_2}{{\bf k}_2^2} \right)_l 
\]
\begin{equation}
\hspace{3cm}{}+({\bf l} \to -{\bf l}).
\label{Tl}
\end{equation}
Here
\begin{equation}
D({\bf k})=\alpha_1(1-\alpha_1)Q^2 + {\bf k}^2 + m_q^2,
\label{10}
\end{equation}
and the function ${\cal F}$ denotes the unintegrated (forward) 
gluon density\footnote{Note that different definitions for ${\cal F}$ exist,
here we use ${\cal F}$ as defined in eq.(\ref{9}).}
which is connected with the usual gluon density $g(x,Q^2)$ through:  
\begin{equation}
\int_0 ^{Q^2}d{\bf l}^2{\cal F}(x,{\bf l}^2) \simeq xg(x,Q^2).
\label{glud}
\end{equation} 
The $\simeq$ sign in the above equation indicates that the relation is 
valid for large $Q^2$. Strictly speaking, the kinematics of diffractive 
$q\bar{q}g$ production requires the nonforward (skewed) gluon density. 
However, 
our cross section formula has been derived in the leading-$\ln W^2$, 
leading-$\ln M^2$ approximation, and the use of the gluon density in 
(12) is valid only in the double logarithmic approximation where 
skewedness is negligeable.       
\par
The parameter $\alpha_1$ is determined by the on-shell conditions for the 
final state particles:
\begin{equation} \alpha_1=\frac{1}{2}\left[ 1+\frac{{\bf k}_1^2}{m^2}-
\frac{({\bf k}_1+{\bf k}_2)^2}{m^2}\pm \sqrt S \right], 
\label{11}
\end{equation}
and it varies between 0 and 1. The values of the
momenta ${\bf k}_1$, ${\bf k}_2$ and of $m^2$ determine the sign in
eq.(\ref{11}). The validity of our cross section formula is restricted
to the kinematic region where the gluon transverse momentum ${\bf k}_2$ is not small. 
\par
The quark mass $m_q$ enters the calculations in two places.
First, the phase space of the diffractive system (and so the parameter
$\alpha_1$ and the function $S$) depend upon the quark mass via the
on-shell conditions for the outgoing particles. Secondly, the
propagators of the internal fermion lines are modified by a nonzero
quark mass which leads to changes in the matrix-elements. Apart from the
function $D({\bf k})$, eq.(\ref{10}), which enters in all four
$\gamma^*p$ - cross sections, an additional term containing the quark mass 
emerges in $d\sigma^{\gamma^*p}_{D,T+}$ (eq.(\ref{2})).  
%%%%%%%%%%%%%%%%%%%%%%%%%%%%%%%%%%%%%%%%%%%%%%%%%%%%%%%%%%%%%%%%%%%%%%%%%%%%%%%%%%%%%%%%%%%%%%
\section{Comparison with measurements}
%%%%%%%%%%%%%%%%%%%%%%%%%%%%%%%%%%%%%%%%%%%%%%%%%%%%%%%%%%%%%%%%%%%%%%%%%%%%%%%%%%%%%%%%%%%%%%
Compared to other charmed particles, $D^{*\pm}$ mesons are easy to reconstruct
which makes them attractive objects for testing diffractive charm production.
$D^{*\pm}$ mesons are identified via the decay channel
\[ D^{*+}\rightarrow D^0\pi^+_{slow} \rightarrow
(K^-\pi^+)\pi^+_{slow} \qquad (\mbox{and c.c.}), \]
which has a branching ratio of 2.63\% \cite{DataG}. In the following comparison
we concentrate on a comparison with the measurement of the 
H1 collaboration~\cite{h1_diff_charm}, who has analyzed data 
collected throughout the years 1995-1997. The amount of
data is still quite poor due to the small branching ratio of the $D^{*}$ meson, 
but higher statistics will come from new data.
\par
We have implemented the cross section formulae for diffractive 
massive $c\bar c$
production~\cite{qqm} and from our expression eq.(\ref{1}-\ref{5}) 
for the massive $q\bar q g$ production into the Monte Carlo generator 
\RAPGAP~\cite{\RAPGAPMC}, which includes full hadronization according to the
Lund string model as implemented in \JETSET~\cite{\JETSETMC}. 
We have used a fixed strong coupling constant
$\alpha_s = 0.25$, and a charm quark mass of $m_q=1.5$~GeV.    
The transition of the charm quark to the $D^*$ meson is performed via the Lund
heavy quark fragmentation function.
Since the experimental measurement does not separate the charge of the 
$D^*$ meson, we have added the cross sections of both charged $D^*$ mesons.
\par
Events generated with \RAPGAP~are selected within the same kinematic region as
in the measurement of H1~\cite{h1_diff_charm}, using electron
and proton momenta (in the HERA system) of 27.6 GeV and 820 GeV, 
respectively, and:
\begin{eqnarray} 
0.05 & \le y  & \le 0.7 \\ 
 2   & \le Q^2 &\le 100 \; \mbox{GeV}^2 \nonumber \\
     & x_{\mathbb P} &<0.04  \nonumber \\
    & |t| & < 1 \; \mbox{GeV}^2 \nonumber 
\end{eqnarray}   
As the measured cross section is integrated over $t$ for 
$|t| \le 1$~GeV$^2$, we have multiplied our cross section 
formulas (eq.(\ref{1}-\ref{5})), valid for $|t|=0$, 
with a phenomenologically motivated $t$
distribution of the form:
\begin{equation}
 f^2(t)=\left(\frac{4-2.8 t}{4-t} 
 \frac{1}{\left(1-\frac{t}{0.7}\right)^2}\right)^2
\end{equation}
The $D^*$ mesons are experimentally detected in the range:
\begin{eqnarray}
|\eta(D^{*\pm})|&< &1.5 \\
p_T(D^{*\pm})& > & 2\;\mbox{GeV} \nonumber
\end{eqnarray}   
with the pseudo-rapidity
$\eta=-\log \tan (\theta /2)$ and the transverse momentum $p_T$ of the 
$D^*$ meson measured in the $ep$ laboratory system.
\par
As can be seen from eqs.(\ref{Til},\ref{Tl}), there is a potential divergency,
if the transverse momentum of the final state gluon ${\bf k}^2_{2}$
approaches zero. 
\begin{figure}[htb]
  \begin{center}
   \includegraphics[width=1\linewidth]{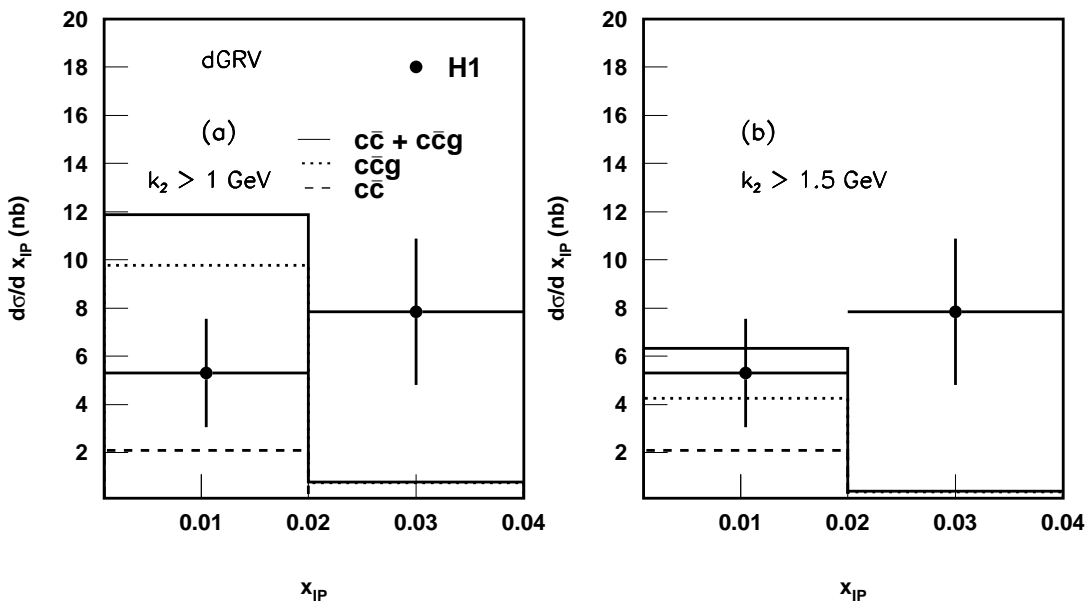}
    \caption{
    The cross section
    $d\sigma/d \log_{10} x_\mathbb P$ for diffractive $D^*$ production 
    within the kinematic
    range specified in the text. The point are the measured cross section from
    H1~\protect\cite{h1_diff_charm}. The prediction obtained with 
    ${\bf k}_{cut}=1\; (1.5)$ GeV is shown in $a$ ($b$). The dashed 
    (dotted) line
    shows the $c\bar c$ ($c{\bar c}g$) contribution alone and the solid line is
    the sum of both. The {\it dGRV} unintegrated gluon density is used, with 
    ${\bf  l}^2_{min} =0.4 \mbox{GeV}^2$.
    \label{fig:grv,xp}}
  \end{center}
\end{figure}
In order to avoid the non-perturbative region, we impose a lower cutoff 
${\bf k}^2_{2cut}$ on the gluon transverse momentum. In our calculations 
we have considered ${\bf k}_{2cut}=1\;\mbox{GeV} $ and 
${\bf k}_{2cut}=1.5 \;\mbox{GeV} $. 
For the unintegrated gluon density, which enters in eq.(\ref{9}), we have used
three different approaches: the derivative of the 
NLO GRV~\cite{GRV} gluon density {\it dGRV},
the unintegrated gluon density $ {\cal F}(x,{\bf  l}^2)$ obtained in the
saturation model of Golec-Biernat and W\"usthoff~\cite{satMod} {\it GBW},
and the CCFM~\cite{\CCFM} unintegrated gluon density 
$ {\cal A}(x,{\bf  l}^2,\bar{q}^2)$ {\it JS} of
\cite{jung_salam_2000,jung-hq-2001}, where $\bar{q}$ defines the evolution
scale, related to the maximum allowed angle of any emission in the angular
ordering approach.
\par
The unintegrated gluon density can be obtained from the integrated gluon
density, if in eq.(\ref{glud})  the $\simeq$ sign is replaced by an 
equality sign:
\begin{equation}
 {\cal F}(x,{\bf  l}^2) = \left. \frac{\partial\;xg(x,\mu^2)}
{\partial\;\mu^2}\right|_{\mu^2 =
  {\bf  l} ^2}
\end{equation}
Here we use for 
$xg(x,\mu^2)$ the NLO GRV~\cite{GRV} gluon density, since it is the only
integrated gluon density available, starting at a low value
of $Q_0^2=0.4$ GeV$^2$. Due to the finite $Q_0$ in any of the available
integrated gluon densities, a lower integration limit
${\bf l}^2_{min} \simeq Q_0^2$ in eq.(\ref{9}) is introduced. Variation 
of this parameter mainly affects the normalization of the cross
sections. For example, when ${\bf l}^2_{min}$ is decreased from $1$ to 
$0.5\;\mbox{GeV}^2$,
the $x_{\mathbb P}$-distribution at ${\bf k}_{2cut}=1
\;\mbox{GeV}$ roughly doubles in the whole $x_{\mathbb P}$ range.
We have chosen to set ${\bf l}^2_{min}$ as small as it is
compatible with the definition of the integrated gluon density, 
so ${\bf  l}^2_{min} =0.4 \; \mbox{GeV}^2$ for {\it dGRV}. 
The other two
unintegrated gluon densities, {\it GBW} and {\it JS}, are defined also for the
very small  ${\bf  l}^2$ region, and therefore no cut needs to be applied there.
In the numerical treatment of the 
$c\bar c$ production cross section neither ${\bf k}^2_{2cut}$ nor 
${\bf l}^2_{min}$ are needed.  
\par
\begin{figure}[htb]
  \begin{center}
   \includegraphics[width=1\linewidth]{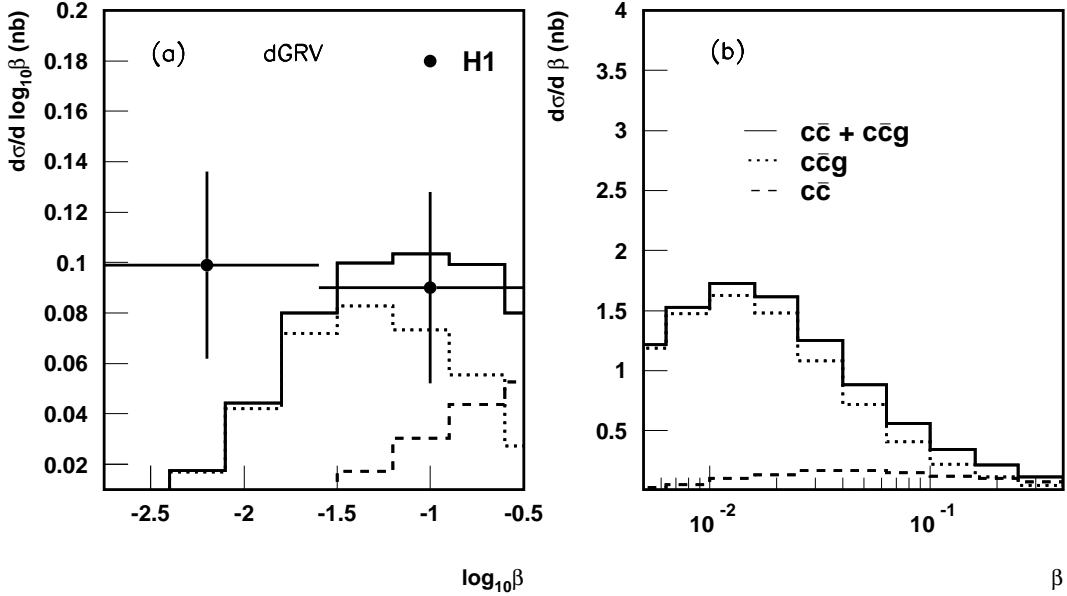}
    \caption{The differential cross section $d\sigma/d \log_{10} \beta$ 
    for diffractive $D^*$ production compared to 
    the measured cross section from H1 $(a)$. 
    In $(b)$  the differential cross section $d\sigma/d \beta$ is shown.
    In all cases we use 
    ${\bf k}_{cut}=1.5$ GeV and the {\it dGRV} unintegrated gluon 
    density. The dashed (dotted) line
    shows the $c\bar c$ ($c{\bar c}g$) contribution alone and the solid line is
    the sum of both. 
    \label{fig:grv}}
  \end{center}
\end{figure}

In Fig.~\ref{fig:grv,xp} we show the effect of changing the cut 
${\bf k}^2_{cut}$ on the differential cross section 
$d \sigma /d log_{10} x_\mathbb P$,
for diffractive $D^*$ meson production in the kinematic region specified above
and compare our prediction to the measurement of H1~\cite{h1_diff_charm}. Also
shown in Fig.~\ref{fig:grv,xp} are the individual contributions of 
$c\bar c$ (dashed histogram) and $c\bar c g$ (dotted histogram).
One clearly sees the reduction of the $c\bar cg$ contribution when the
cutoff in ${\bf k}_{cut}$ is increased. Since our calculation is valid 
for small $x_\mathbb P$, we focus only on agreement in the 
low $x_\mathbb P$ region. A cutoff value of 
${\bf k}^2_{cut}=1.5 \;\mbox{GeV}^2$ seems to be a reasonable choice 
which we will keep independently of the choice of the unintegrated gluon
density.
With this cut the computed cross section in the lower 
$x_\mathbb P$ bin agrees well with the data. In the upper $x_\mathbb P$ bin, 
however, the theoretical curve is by a factor of about 10 smaller than the 
data point (ignoring the large error on the measurement). 
In this $x_\mathbb P$-region it is expected that, apart from Pomeron 
exchange (which, in our model, is the 2-gluon exchange) 
secondary exchanges have to be included: in a perturbative description 
such an exchange corresponds to $q\bar{q}$-exchange. Since such a contribution
has not yet been included into our calculation, it is not surprising 
that the two-gluon model undershoots the data.
\par
In Fig.~\ref{fig:grv}$a$ we compare our prediction of
the cross section for diffractive $D^*$  production  as a function of 
$\log_{10} \beta$ with the measurement of H1 
(it is understood, that all other variables, 
in particular $x_\mathbb P$, are integrated over).
We also show the individual contributions of 
$c\bar c$ (dashed histogram) and $c\bar c g$ (dotted histogram).
Clearly, the $c\bar{c}g$ is badly needed in order to get closer
to the data than with $c\bar{c}$ alone.
The measured cross section is slowly rising with decreasing 
$\log_{10} \beta$. The theoretical curve does not quite follow this rise,
the reason for this is the correlation between $\beta$ and $x_\mathbb P$: 
small $\beta$ values require large diffractive masses $M$, which due to the
kinematic restrictions in the analysis, are predominantly produced at large   
$x_\mathbb P$. As argued above the large   
$x_\mathbb P$-region  needs secondary exchange which in our approach is not yet
included.  
For illustration, we show in Fig.~\ref{fig:grv}$b$ 
the cross section for diffractive $D^*$  production $d\sigma / d \beta$. We
observe that the strong drop in cross section at small $\beta$, as seen in 
Fig.~\ref{fig:grv}$a$ is a consequence of plotting 
$d\sigma / d \log_{10}\beta$ instead of $d\sigma / d \beta$. From 
Fig.~\ref{fig:grv}$b$ we see that the theoretical curve increases towards low
$\beta$, and only at $\beta < 0.01$ is decreasing, which again is a consequence
of kinematic correlations.
If the double differential 
cross section $d^2 \sigma /d\beta dx_\mathbb P$ is considered, 
our prediction shows the expected rise towards small $\beta$ at fixed 
$x_\mathbb P$.    
The shape of the theoretical curves as a function of $\beta$ 
is almost independent of the choice of 
the cutoffs ${\bf l}^2_{min}$ and ${\bf k}_{cut}$, as they mainly affect the 
overall magnitude of the cross section. 
In the small $\beta$ region, corresponding to large diffractive masses $M$, also
the radiation of more than one gluon, such as $c {\bar c} g g$ need to be
considered. Since experimentally the cross section is defined as 
$ e+p \to e ' + (D^* + X) + p_{diff}$, 
where in the diffractive system $M= D^* + X$, the hadronic state $X$ 
is not further
specified or measured, multiple soft gluon contributions might be present in the
data, which have yet not been estimated consistently in the perturbative
calculations.
\par
\begin{figure}[htb]
  \begin{center}
   \includegraphics[width=1\linewidth]{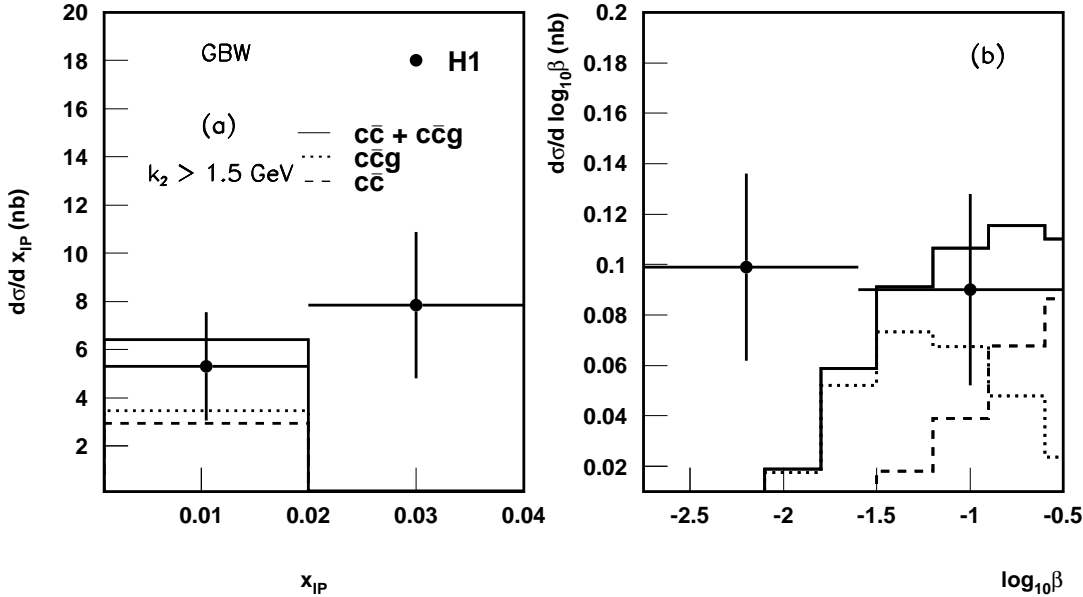}
    \caption{The differential cross section $d\sigma/d \log_{10} \beta$ 
    for diffractive $D^*$ production compared to 
    the measured cross section from H1 $(a)$. 
    In $(b)$  the differential cross section $d\sigma/d \beta$ is shown.
    In all cases we use 
    ${\bf k}_{cut}=1.5$ GeV and the {\it GBW} unintegrated gluon 
    density from the saturation model.
    The dashed (dotted) line
    shows the $c\bar c$ ($c{\bar c}g$) contribution alone and the solid line is
    the sum of both.}
    \label{fig:sat}
  \end{center}
\end{figure}
The saturation model of K.Golec-Biernat and M.W\"usthoff \cite{satMod} describes
a completely different approach to estimate the unintegrated gluon density 
${\cal F}(x,{\bf l}^2)$, which is needed in eq.(\ref{9}).
In this model the total $\gamma^*p$ cross section is 
described by the interaction of a $q\bar{q}$ pair (dipole) with the proton, 
and a particular ansatz is made for the dipole cross section. The
function ${\cal F}(x, {\bf l}^2)$ has the form
\begin{equation} 
 {\cal F}(x, {\bf l}^2) = \frac{3 \sigma_0}{4 \pi^2 \alpha_s } \, R^2_0(x)
{\bf l}^2 e^{-R_0^2(x)\;{\bf l}^2} \quad ,\qquad
R_0=\frac{1}{\mbox{GeV}} \left (\frac{x}{x_0} \right)^{\lambda /2},
\label{sat}
\end{equation} 
and the three parameters of the model are determined by fitting
inclusive DIS data
(including charm with: $\sigma_0=29.12$ mb,$\,\lambda=0.277$,
$\,x_0=0.41\;10^{-4}$ and $\alpha_s=0.2$  
\cite{satMod}). For large ${\bf l}^2$, ${\cal F}(x, {\bf l}^2)$ 
has the meaning of the unintegrated gluon density, but for smaller 
${\bf l}^2$ the function looses this interpretation and has to be viewed as 
a (model dependent) extrapolation. The ansatz eq.(\ref{sat}) 
holds for the $q\bar{q}$ 
color dipole cross section. The $q \bar{q}g$ system consists of a color 
triplet, and anti-triplet and a color octet, and one expects that the 
dominant configuration is a dipole consisting of two octets: in eq.(\ref{sat})
we therefore rescale the color charge and use:
\begin{equation}
{\cal F}(x, {\bf l}^2) = \frac{3 \sigma_0}{2.25^2 \cdot 4 \pi^2 \alpha_s } 
\, R^2_0(x)
{\bf l}^2 e^{ \frac{ -R_0^2(x)}{2.25}\;{\bf l}^2} \quad .
\label{sat1}    
\end{equation}  
Insertion of the ansatz (eqs.(\ref{sat},\ref{sat1})) into eq.(\ref{glud})
leads to an integrand that vanishes as ${\bf l}^2$ goes to zero. 
Therefore, within this model we no longer  need 
any lower cutoff in the ${\bf l}^2$ integral, and our calculation 
provides absolute predictions of the cross sections (note, however, that 
we still have the cutoff ${\bf k}_{cut}=1.5
\;\mbox{GeV}$ on the final state gluon). 
Fig.~\ref{fig:sat} shows the differential cross section 
$d \sigma / d \log_{10} x_{\mathbb P}$ calculated using 
eq.(\ref{sat},\ref{sat1}) compared
to the measurement of H1.
The calculated cross section ($c\bar c$+$c\bar cg$)
is similar to that of Fig.~\ref{fig:grv}, and the same discussion applies.
Also the $\beta$ distribution (Fig.~\ref{fig:sat}$b$) is very similar to 
the previous model. Note that without the correcting color factor in the  
dipole cross section formula the $c\bar{c}g$ cross section would be 
larger by about a factor of about 1.5. 
\begin{figure}[htb]
  \begin{center}
   \includegraphics[width=1\linewidth]{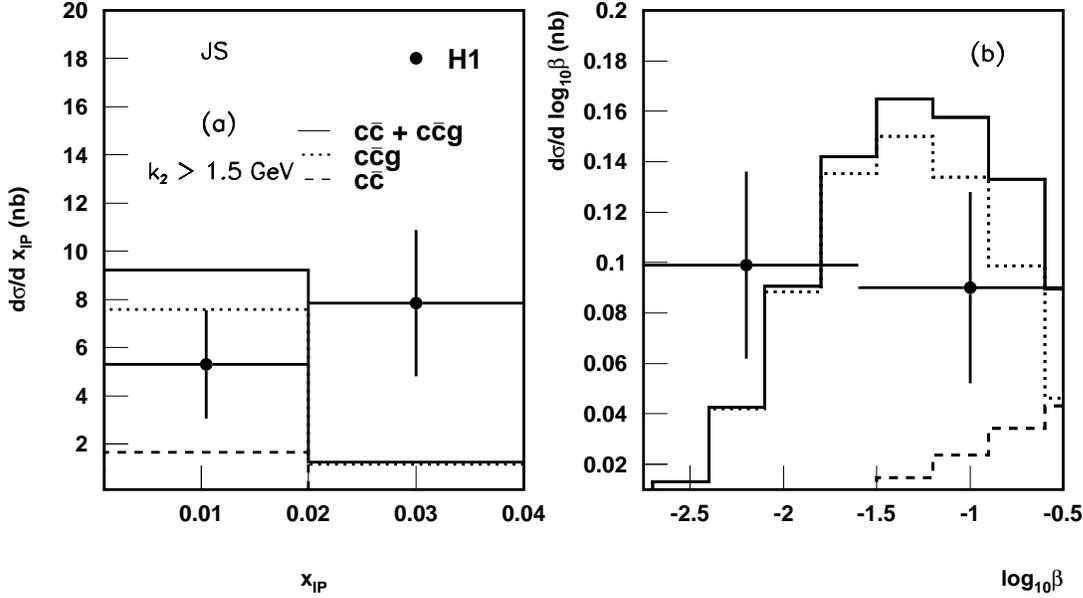}
    \caption{
    The differential cross section $d\sigma/d \log_{10} \beta$ 
    for diffractive $D^*$ production compared to 
    the measured cross section from H1 $(a)$. 
    In $(b)$  the differential cross section $d\sigma/d \beta$ is shown.
    In all cases we use 
    ${\bf k}_{cut}=1.5$ GeV and the CCFM based {\it JS} unintegrated gluon 
    density. The dashed (dotted) line
    shows the $c\bar c$ ($c{\bar c}g$) contribution alone and the solid line is
    the sum of both.}    \label{fig:ccfm}
  \end{center}
\end{figure}
\par
The unintegrated gluon density, based on the
consistent treatment of color coherence effects, is described by the CCFM
evolution equation~\cite{\CCFM}. According to the CCFM equation, the emission 
of partons during the initial state cascade is allowed only in an 
angular-ordered
region of phase space. In the large (small) $x$ limit, the CCFM equation is
equivalent to the DGLAP (BFKL) evolution equations, respectively. A
solution of the CCFM equation has been found, which successfully can be used to
describe a bulk of measurements at HERA and the Tevatron 
\cite{jung_salam_2000,jung-hq-2001,jung_ringberg2001}.
However, due to the angular ordering requirement, the unintegrated gluon 
density ${\cal A}(x, {\bf l}^2, \bar{q}^2)$
is now also a function of the evolution scale $\bar{q}$, which is related to the
maximum allowed angle. Here, this scale is set either by the $q \bar{q}$ pair,
or by the final state gluon for $q \bar{q} g$:
\begin{equation}
\bar{q }^2  = \left\{ \begin{array}{ll} 
             m_{q \bar{q}}^2 + {\bf Q}_t^2 & \mbox{for}\;\; q\bar{q} \\
             \frac{{\bf k}_2^2}{1-z} & \mbox{for}\;\; q\bar{q} g
                    \end{array} \right.
\end{equation}
with ${\bf Q}_t$ being the vectorial sum of the transverse momenta of the 
$q\bar{q}$ pair, 
and $z=(Q^2+m_{qq}^2)/(Q^2 +M^2)$.
Since the explicit parameterization of ${\cal A}(x, {\bf l}^2, \bar{q}^2)$ from
\cite{jung_salam_2000} is valid also in the very small ${\bf l}^2 $ region, no
cut on ${\bf l}^2 $ needs to be applied for the integral in eq.~(\ref{9}). The
results for the 
differential cross sections as a function of
$x_{\mathbb P}$ and the $\log_{10} \beta$  shown in
Fig.~\ref{fig:ccfm}$a$ and $b$ are quite similar
to those of the unintegrated gluon density from 
saturation model {\it GBW} and or from the derivative of the integrated gluon
density {\it dGRV}. 
The main difference is the ratio of
$c\bar c$ and  $c{\bar c}g$ contribution.
The enhancement of the $\beta$ distribution by
$c{\bar c}g$ is much stronger than in case of the two other models. 
%%%%%%%%%%%%%%%%%%%%%%%%%%%%%%%%%%%%%%%%%%%%%%%%%%%%%%%%%%%%%%%%%%%%%%%%%%%%%%%%%%%%%%%%%%%%%%
\section{Conclusion}
%%%%%%%%%%%%%%%%%%%%%%%%%%%%%%%%%%%%%%%%%%%%%%%%%%%%%%%%%%%%%%%%%%%%%%%%%%%%%%%%%%%%%%%%%%%%%%
In this article we have analyzed DIS diffractive charm production 
(production of $D^{*\pm}$ mesons) within the perturbative two-gluon model.
For the two-gluon amplitude we have used three different models: 
the unintegrated gluon density derived from the integrated
DGLAP gluon density
{\it dGRV},
the saturation model of Golec-Biernat and W\"usthoff {\it GBW},
and a CCFM-based unintegrated gluon density {\it JS}
(the last two models are parameter-free, the first one 
depends upon a cutoff on the internal momentum integral).   
In all three cases the calculated cross sections are of the same order of 
magnitude as the data, and within the kinematical region where the models
apply, the shapes of the cross sections are consistent with the data. 
Compared to an earlier attempt (with the {\it dGRV} gluon density) where
only $c\bar{c}$ production had been included in the theoretical analysis
the present analysis contains, as the new ingredient, also (massive)
$c\bar{c}g$ production and leads to a considerable improvement in the 
agreement with experimental data.
\par 
It is encouraging to see that, for the {\it dGRV} gluon density, 
our analysis of diffractive charm production uses the same parameters 
as in the successful analysis of diffractive jet production and we were able to
consistently describe both types of processes.
\par
We view the use of the two-gluon model as part of a more general strategy 
of analyzing DIS diffraction data at HERA. 
In a first step one would analyze those 
diffractive final states which are dominated by short distances 
(diffractive jets or states consisting of heavy quarks): in these 
processes the application of the two-gluon model can be justified.
In a second step, one would try  
to extrapolate also into kinematic regions where soft physics becomes 
important. In the present analysis, such extrapolations are contained 
in the saturation model and in the CCFM amplitude; the use of the GRV 
gluon density, on the other hand, requires a momentum cutoff.
In a final (and future) step one would need to find a QCD-based `derivation' of 
the extrapolation from hard to soft physics.
\par 
Despite this encouraging success, several improvements in the theoretical
part of our model should be made. First, since the cross section formula for     
$c\bar{c}g$ production has only been calculated in the 
leading log-$M^2$ approximation, an improvement which extends the applicability
down to small-$M^2$ values would be very desirable. For consistency reasons,
one then would need a NLO-calculation of $c\bar{c}$ production. 
Results of such a calculation would also allow to eliminate the cut on the
transverse momentum of the final state gluon.
Next, since 
the region of $x_\mathbb P>0.02$ seems to require secondary exchanges, they
should be  modeled, in the framework of perturbative QCD,  by $q\bar{q}$
exchange. Finally, our comparison with data indicates the need of 
$c\bar{c}gg$ final states: 
such an extension (at least in the leading log-$M^2$ approximation)
should be fairly straightforward. 
A successful test of the two-gluon model in DIS Diffraction, apart form 
providing a description of charm or jet production at HERA, is also of 
general theoretical interest: the cross section formula for diffractive 
$q\bar{q}+ng$ production contains the perturbative triple Pomeron vertex 
which is expected to play a vital role in the unitarization of the BFKL
approximation. 
It has been calculated both analytically and numerically, and these
calculations can be tested experimentally in DIS diffraction
dissociation.   
\raggedright

\end{document}